# Tongue pressure recordings during speech using complete denture

Christophe Jeannin[*1,5], Pascal Perrier[1], Yohan Payan[2], André Dittmar[3], Brigitte Grosgogeat[4,5]

[1]Institut de la Communication Parlée, INPG, Grenoble, France

[2] TIMC, Grenoble, France

[3] Laboratoire des microsystèmes et microcapteurs biomédicaux, INSA, Lyon, France

[4] Laboratoire d'Etude des Interfaces et des biofilms en Odontologie, Université Lyon I, France

[5] Hospices Civils de Lyon, Service d'odontologie, Lyon, France

**This paper describes an original experimental procedure to measure mechanical interactions between tongue and teeth during speech production. Using edentulous people as subjects, pressure transducers are inserted in their complete denture duplicate. Physiology is respected during sound and pressure recording as with standard complete denture. Original calibration device is also described in order to know what kind of information can be extracted from the data. Then the first results of the pilot experiments are presented.**

---

Corresponding author : Christophe JEANNIN, INPG laboratoire Gipsa-lab (ex ICP), 46, avenue Félix Viallet 38031 Grenoble cedex 1 – France -

E-mail: jeannin@icp.inpg.fr

Phone and fax: 33-476 461 723

# I. INTRODUCTION

Clinical observations about the interactions between tongue and complete denture, especially during speech are reported since 70 years [1]. In order to formalize these empirical data, experimental measurements in the mouth are necessary. Therefore devices for measurement are needed.

The existing devices don't preserve physiological conditions during recordings, because of their position.

The aim of this paper is to present an original device for tongue pressure measurement usable only for the edentulous people but which creates no change in the mouth during the recordings.

This device should allow realizing a biomechanical map of tongue pressures during speech between tongue, teeth and palate. Moreover, according to the literature [2,3], this new experimental procedure aims at:

(1) measuring the interaction between tongue, teeth and palate without perturbing the production of speech, and

(2) studying how speech motor control strategies evolve for edentulous people, from the moment when an artificial denture is put back into the mouth

# II. DESCRIPTION AND METHODS

The basic principle and the originality of the method presented are to use edentulous people as subjects. Pressure sensors are placed in their complete denture, in such a way that the geometry of the vocal tract remains exactly the same during the recordings or not.

## A. *Description*

<u>Background</u>

There is no study about transducers included in a prosthodontic device. Pressure measurements are realised in dentate subjects. Therefore, sensors are stuck on the teeth [4] or included in an artificial palate fixed on the teeth [5,6,7]. In these conditions, physiologic situation can not be considered like usual for speech production.

Different kinds of sensors have been tested in order to compare their efficiency and their repetitive abilities [8]. However, comparisons between transducers' locations have been realized to assess less disturbing positions for speech production [2].

The strain gauge sensor, described in this paper was chosen because of its thermic stability and its mechanical properties [9]. It was already used in previous studies, using dentate subjects, for tongue pressure measurement [4]. According to its calibration device, it allows a good description of the viscoelastic structure of the tongue.

<u>Description of the complete denture</u>

According to Gysi's theory rules of complete denture making for molar and premolar positions, the artificial palate is 2,5 mm thick [1]. Hence, the pressure sensor and the wires can both be easily inserted into the complete denture, without creating any additional change of the oral cavity in relation to the usual complete denture (fig.1).

For each edentulous patient, the dental prosthesis designed for obvious medical purposes is accurately reproduced thanks to the technique of duplication. Hence 5 duplicates are made in order to have different possible positions for the pressure sensor. For each duplicate, a different position is determined for the sensor, using palatogram recordings (see Method) . As the sensors are inserted before the experiment, no time is wasted between each recording. Therefore, as tongue interactions are tested at different locations, disturbances between the recordings are minimized.

The duplicate can support several transducers, but currently the amplifier can accept only two of them.

<u>Description of the experimental device</u>

The pressure sensor is included in the complete denture. According to the literature, wires go to the external connector from the vestibular side of premolar area by the way of labial commissure. Then, a wire goes from the connector to the amplifier, next to a data sampling board and at last to the computer.

A microphone is also connected to an analogic amplifier to record the acoustic speech signal simultaneously with the pressure signal. This amplifier is in turn linked to the same data sampling board.

<u>Transducer description</u>

It’s made of a strain gauge sensor from whence goes a insulating sheath to a connector. There are 5 wires made of copper which are 6/100 diameter, 4 for the strain gauge and one for the ground.
The connector is used to connect the transducer to the amplifier outside the mouth; it is hitched onto a clothespin to preclude wire traction.

Strain gauge sensor description

The intraoral area is a very difficult environment mainly because of three factors:

- Permanent moisture due to saliva
- Variable temperature
- Mechanical constraints

Therefore, the sensor must be electrically insulated and water proof. Moreover, it has to resist to many experiments, so it has to be sturdy.

The strain gauge sensor is composed of thirteen layers (fig. 2). Each layer plays an important part in measuring the abilities of the transducer [4].

The transducers, including the sensor are handmade in the ICP laboratory.

The middle layer is made of a steel cantilever beam which is 10/100 mm thick. It supports on each side an active strain gauge. The gauges (Vishay, ref EA06 062 AQ 350) are placed in a half Wheatstone bridge configuration. This strain gauge has been chosen because of its stability in temperature [8].

Gauges are bonded on a metallic support with M-Bond 200 adhesive (Vishay measurement group). Wires of 0,1mm diameter are soldered with tin on the gauge. The solder area is $1mm^2$ small and there are 4 points, 2 by side.

The cantilever beam is chemically prepared in order to bond the gauges (etching, rinsing, neutralizing with sandpaper size 400). The three shallow layers are made of protective coating called M-Coat A (Vishay measurement group) for electric insulation and the last one is made of a dental varnish called Copalite®.

In any case, during the construction, the thickness of all the liquid components like protective coating or bonding must be as small as possible in order to preserve the mechanical properties of the sensor. Consequently, just one application of each component by layer must be done.

Figure 3 shows a detailed picture of such sensor.

Associated instrumentation tools:

Before sampling, the pressure signal goes to an analogic amplifier (2100 series by Vishay Measurement) which is made of 3 modules:

- A power supply module (2110B) for bridge excitation which provides a 2V current for this experiment.
- A strain gage conditionner (2120B). It includes bridge completion, bridge balance, amplifier balance, bridge excitation regulator and shunt calibration.
- A digital display (2130) that provides real-time digital readout.

To record the acoustic signal, a microphone is placed near the patient. The acoustic signal goes to a dedicated analogic amplifier (realized in the ICP laboratory) and then to the data sampling board (DT 9800 series by Data translation) that is connected directly to the host computer via an USB port.

This board can accept up to 16 analog inputs that can be simultaneously sampled at different rates (from 50 Hz to 20 000 Hz). For these experiments, until 3 inputs are used: one for sound and one or two for the tongue pressure.

*B. Method*

20 patients are selected, 10 of them are new complete denture wearers and the 10 others have already a complete denture which has to be replaced. All of them, male or female have French as native language and are older than 18. They mustn’t be deaf or in treatment for this pathology and mustn’t have any difficulties of comprehension.

Duplicates and palatograms

At the end of the prosthodontic treatment, duplicates are realized at the same time as the complete denture’s polymerization.

In order to know where the sensors to measure tongue-palate/teeth interactions during speech production have to be inserted, the exact locations of tongue contacts are determined [11]. Therefore, in a preliminary stage, palatograms are carried out, with the duplicates:

Indeed, electropalatagraphy (EPG) cannot be used with complete denture because of obvious incompatibilities (weight of EPG device and prosthodontic retention) [12].

Therefore, pink powder occlusion spray ("okklufine premium™") is applied on the teeth and on the artificial palate. When the tongue is in contact with the powder, the powder is removed from the contact area.

Thus, when the subject pronounces a specific phoneme ("T" sound, for example), the edges of the contact areas are determined. Then, these edges are highlighted with a black pen before the powder is removed (fig 4).

This technique is not as accurate as EPG, but enough accurate to determine where the sensors have to be set. Then the transducers are included in the marked areas.

Pressure sensor calibration

The transducers are handmade and their situation are different for each of them in the duplicate. So, they have to be calibrated individually to convert electric signal into mechanical units such as strength or pressure. Figure 5 shows a schematic drawing of this calibration device.

The soft body structure of the tongue suggests that pressure should be more appropriate to convert electric signal. Indeed, due to this mechanical characteristic, interaction between tongue and external structures is more a surface one than a point to point contact.

However, the visco-elastic properties of the tongue make measurement of this contact pressure difficult. Indeed, the shape of the tongue varies over time in presence of contacts, and this variation is strongly dependent on the tongue elastic properties. Hence, formalizing the relation between the strain exerted on the sensor and the contact pressure is not a simple task. So an original device called "dried water column" has to be invented:

The weight of a water column is applied to the whole surface of the sensor via a latex membrane. The shape of this membrane is specific and changeable (fig 6). It can be freely deformed by the weight of the water, but has no stiffness. It means that the electric signal is only due to the water pressure. Indeed, as there is no water in the column, there is no pressure on the sensor, so the electric signal is zero.

This is a nice way to model the soft body structure of the tongue, which give informations about pressure when there's contact.

This makes possible the calibration of the sensor placed in the duplicate. The prosthodontic device is blocked with an adjustable screw. Then, using perpendicular slides, it's adjusted at right angle between the sensor and the water column. Finally, the position in height of the water column and the latex membrane are set up until the latex membrane touches lightly the sensor (0V for 0cm). Then, the upper tank is moved, using different wedges, successively at different heights. The electric signal and the heights are recorded and written down. The calibration chart (fig 7) shows different heights where the upper tank has been moved.

The calibration of the sensor has to be done very carefully and to be explained in order to know what kind of information can be extracted from the data. Figure 7 shows the linear relation between transducer output and height of water using this calibration device.

Measurements

Sounds are repeated several times by the subjects in isolation and using tongue-twisters such as, for the alveolar stop /t/, French syllable /ta/ and French short sentence: /totoatetesatetin/.

Amplifier, microphone, data sampling board and computer are set. During the recordings, the patient is sitting on a dental chair placed in a right position. The microphone is placed at 25 cm in front of him. The tongue pressure and the sound are recorded simultaneously.

Moreover, measurements are realized in different experimental conditions:

- Measurement device as previously described
- Measurement device as previously described + noised headphone into which miscellaneous sound frequencies are sent to isolate the patient from the outside, thus removing the external auditive feed-back.
- Measurement device as previously described + tongue anesthesia which is administered to desensitize the tip of the tongue (made with a topical anesthetic gel, Topex®) in order to remove the sensitive effect of the tongue during pressure on the palate.
- Measurement device as previously described + noised headphone and tongue anesthesia

These should improve the knowledge of the part of these phenomena in the tongue control adaptation during speech.

These recordings are repeated during 6 weeks to control the value of tongue pressure in the same areas and to show potential new interaction areas between tongue and palate.

Finally, the data are analyzed; the comparison is made during the time of experimentation for the same patient, each of them is his self check-sample.

## III. RESULTS and DISCUSSION

### *A. Preliminary results*

The data are exported in matlab® software and then are extracted in different files for sound and pressure and for different experimental conditions.

The sound files are analyzed and labeled with Praat® software in different stages including the burst. Initially, the burst onset, which characterizes the moment at which the airflow abruptly increases after the consonantal vocal tract closure has been released, is only considered.

Then the acoustic and the pressure signals are observed in parallel for each experimental condition, in order to observe the potential influence of the experimental condition on their synchronization.

Figure 8 shows an example of the preliminary results obtained for /ta/ sound in a pilot study carried out with a 72 year old female subject. It describes the results obtained during one recording session. The sensor was inserted in the most front contact area measured with palatography, it means on the palate behind the teeth.
It can be seen that the acoustic release is well-synchronized with the abrupt decrease of tongue pressure in the palate, which corresponds to the acoustic data for the /t/ sound

Figure 9 shows that despite the differences observed in the different experimental conditions, it seems that the duration of the offset is quite constant and its value is close to 30ms. It is assumed that this interval could correspond to the time that is physically required for the airflow to sufficiently increase after the closure release, in order to generate the turbulent acoustic source associated with stop production. The physical constraint could namely explain why this duration is not modified with the experimental condition.

Figure 10 shows the maximal tongue/teeth pressure intensity measured during the /t/ closure in repeated /ta/ sequences for the different experimental conditions. It suggests that there is an effect due to the auditory perturbation associated with the noise (pink and light blue points versus yellow and dark blue points). For the normal and for the anesthesia conditions, the maximal intensity is smaller in auditory perturbed conditions than in their unperturbed counterpart. This effect is observed only on the intensity of the signal, whereas the duration remains steady.

As parts of the pilot study, these previous results (fig 9&10) have to be compared to the next ones, in order to know if the trends observed are confirmed for the same subject for the different experimental conditions during time.

### *B. Discussion*

The main point of this study is to admit that complete denture wearers are in physiological conditions using their prosthodontic device. Indeed, the patients eat, speak and sleep with their prosthodontic devices; they have to remove it only for washing. In this case, the complete denture can be considered as a part of the physiological structures.

But, the idea of physiological versus pathological conditions can be easily discussed, such as invasive or not invasive device. Indeed, a complete denture has got an artificial palate, gum and teeth. It is from 2 to 3 millimeters thick. Its measurements are around 7 cm to 7 cm, so it can be considered as an invasive system.

Can complete denture be really physiological? However, edentulous people are not in physiological conditions. The vertical dimension of occlusion, which refers to the degree of separation between the mandible, or lower jaw-bone and the maxillae, or upper jaw-bone is often altered. [13]

Indeed, the alveolar bone is removed with the loss of teeth. In addition to these losses, there are no more structures left to stop the vertical dimension, the mandible can over go, the tongue uses all the free space between the jaws, the lips are not held up by the teeth. Moreover, according to this situation of altered vertical dimension, temporomandibular disorders (with or without symptoms) exist. The condition of the mouth has become pathologic.

Then, complete denture is a prosthodontic treatment to restore physiological conditions by replacing the missing spaces (alveolar bone, teeth, gum) and restoring the altered functions (mastication, speech, aesthetic).

Therefore, despite its size, complete denture cannot be considered as invasive because it is integrated in the mouth where it replaces such missing structures.

Consequently, the duplicates used for tongue pressure measurement can be rated, like complete denture, as not invasive device.

## IV. CONCLUSION

An original device for the measurement of the mechanical interactions between tongue and teeth and/or palate is presented. It is adapted to a specific kind of subjects, namely edentulous patients. Using the complete denture to insert sensors, the device permits the measurement of contact pressure without introducing any additional perturbation other than the prosthesis itself.

This experimental setup permits to study speech production either by patients who have worn their prosthesis for years and have completed the adaptation process to it, or by patients who have just received the prosthesis, in order to study how they adapt to their new denture.

The preliminary results show a well synchronization between the 2 signals. The differences between the different experimental conditions should be verified with the next results. Several hypotheses are emitted about the results and the mechanical process involved in the speech production.

## ACKNOWLEDGMENTS

This study was supported by the EMERGENCE research program of the Région Rhône Alpes and is supported for the clinical experiments by the Hospices Civils de Lyon. It has been approved by a French CCPPRB committee and has a Clinical Trials Identifier (NCT 00273078).

## FIGURE CAPTIONS

**Figure 1: Experimental device**

**Figure 2: description of the sensor**

**Figure 3: overall view of the transducer**

**Figure 4: highlighted areas of tongue contacts during \ta\ sound after palatogram**

**Figure 5: description of the calibration device**

**Figure 6 : Shapes of the latex membrane with different heights of water**

**Figure 7: Linear relation between height of water and transducer output during calibration**

**Figure 8: Acoustic signal (in blue) and pressure signal (in red) for the /ta/ sound with the sensor alone**

**Figure 9 : Time interval (mean value and variance) from the tongue/teeth pressure offset to the consonant burst onset for different experimental conditions**

**Figure 10: Effects of the different experimental conditions onto the tongue/teeth maximal pressure intensity**

FIGURES

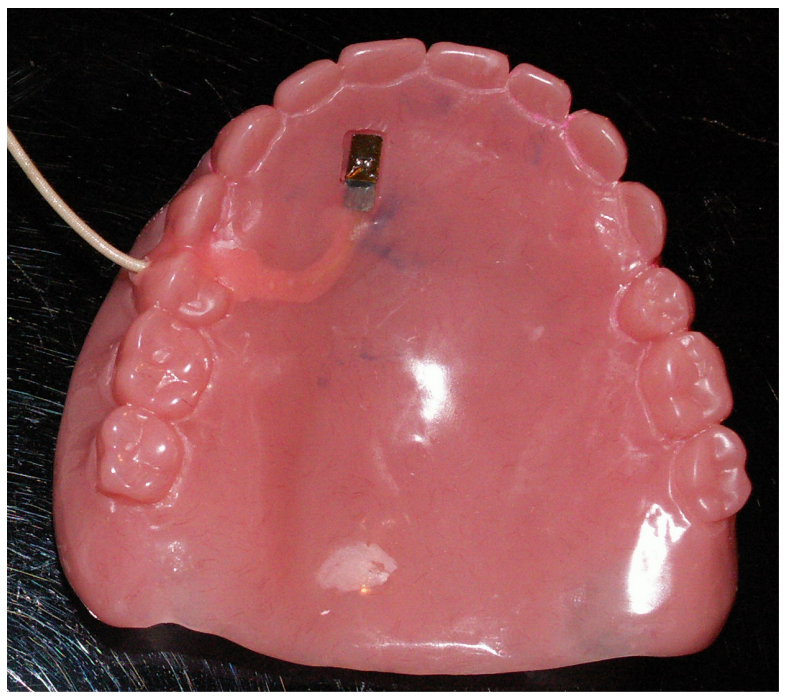

**Figure 1**

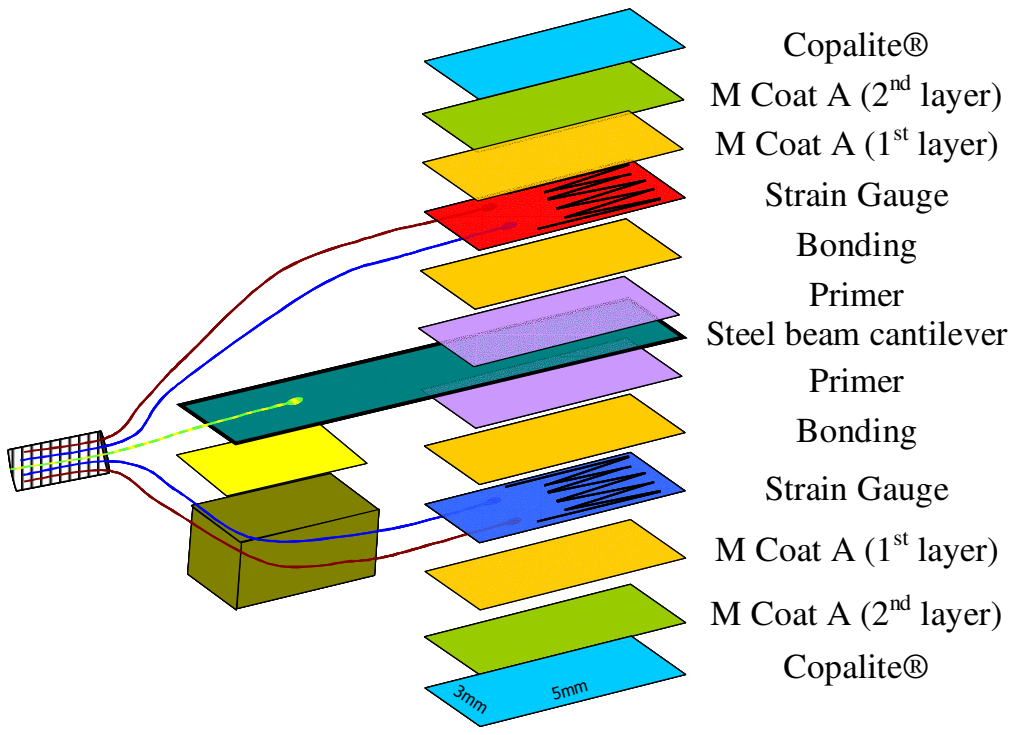

**Figure 2**

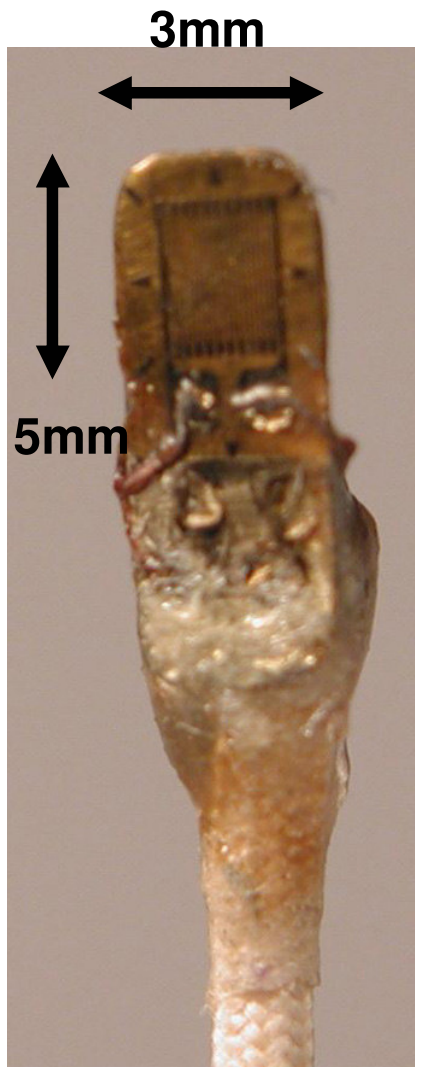

**Figure 3**

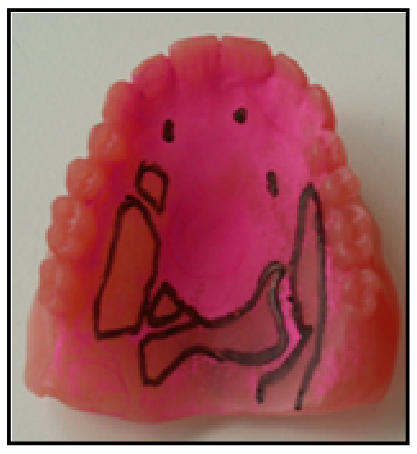

**Figure 4**

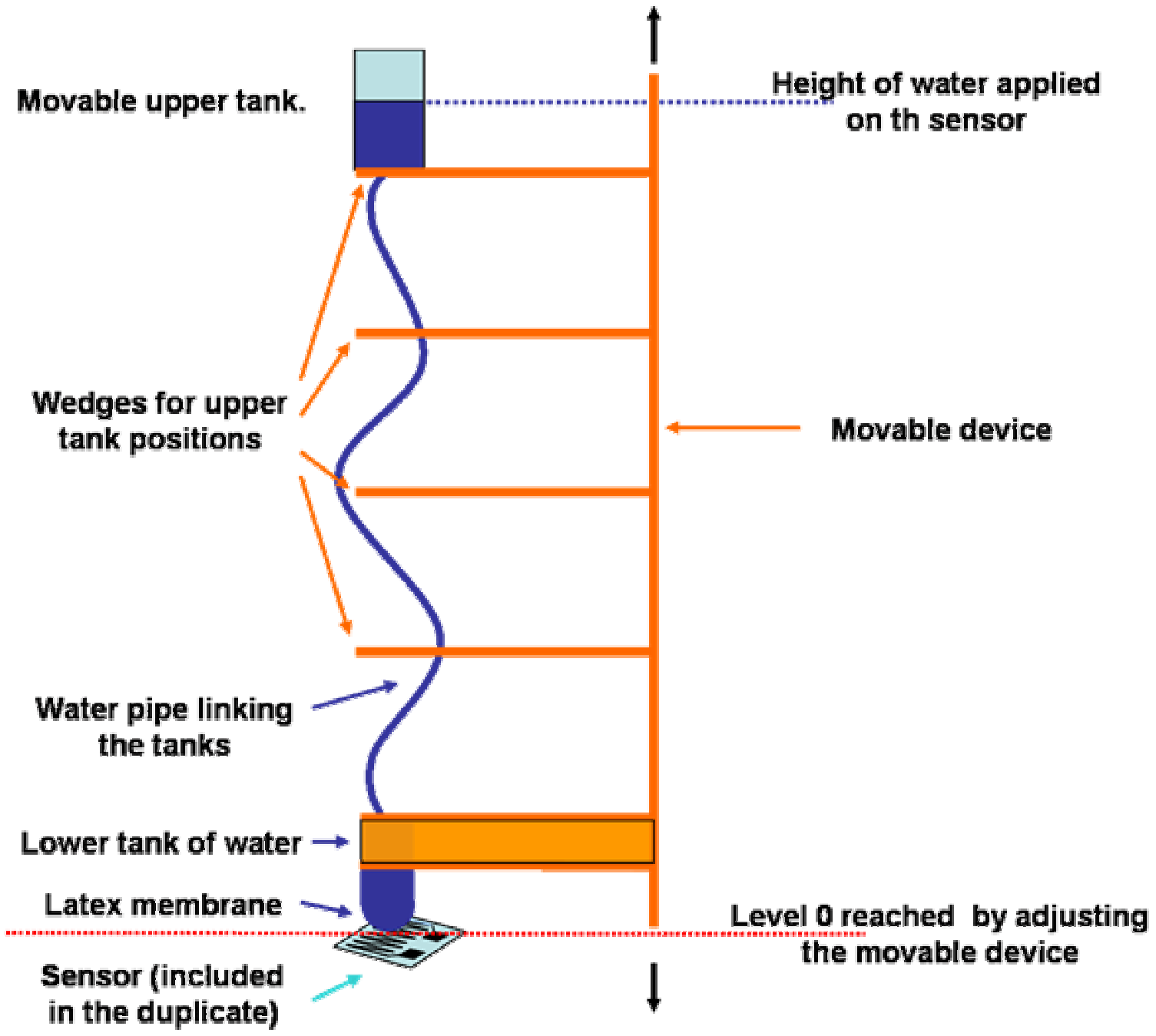

**Figure 5**

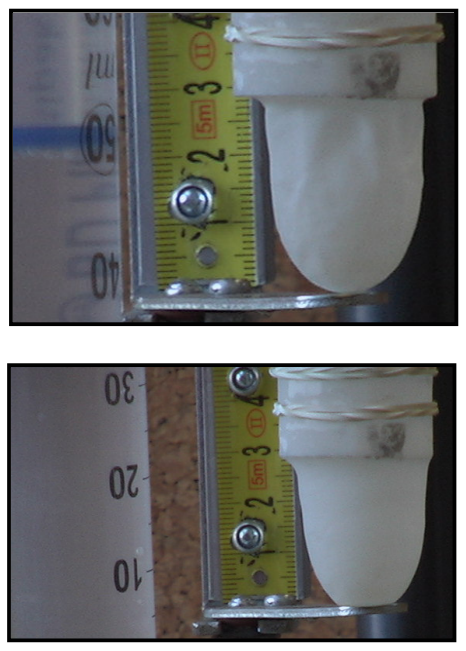

**Figure 6**

## CHARTS

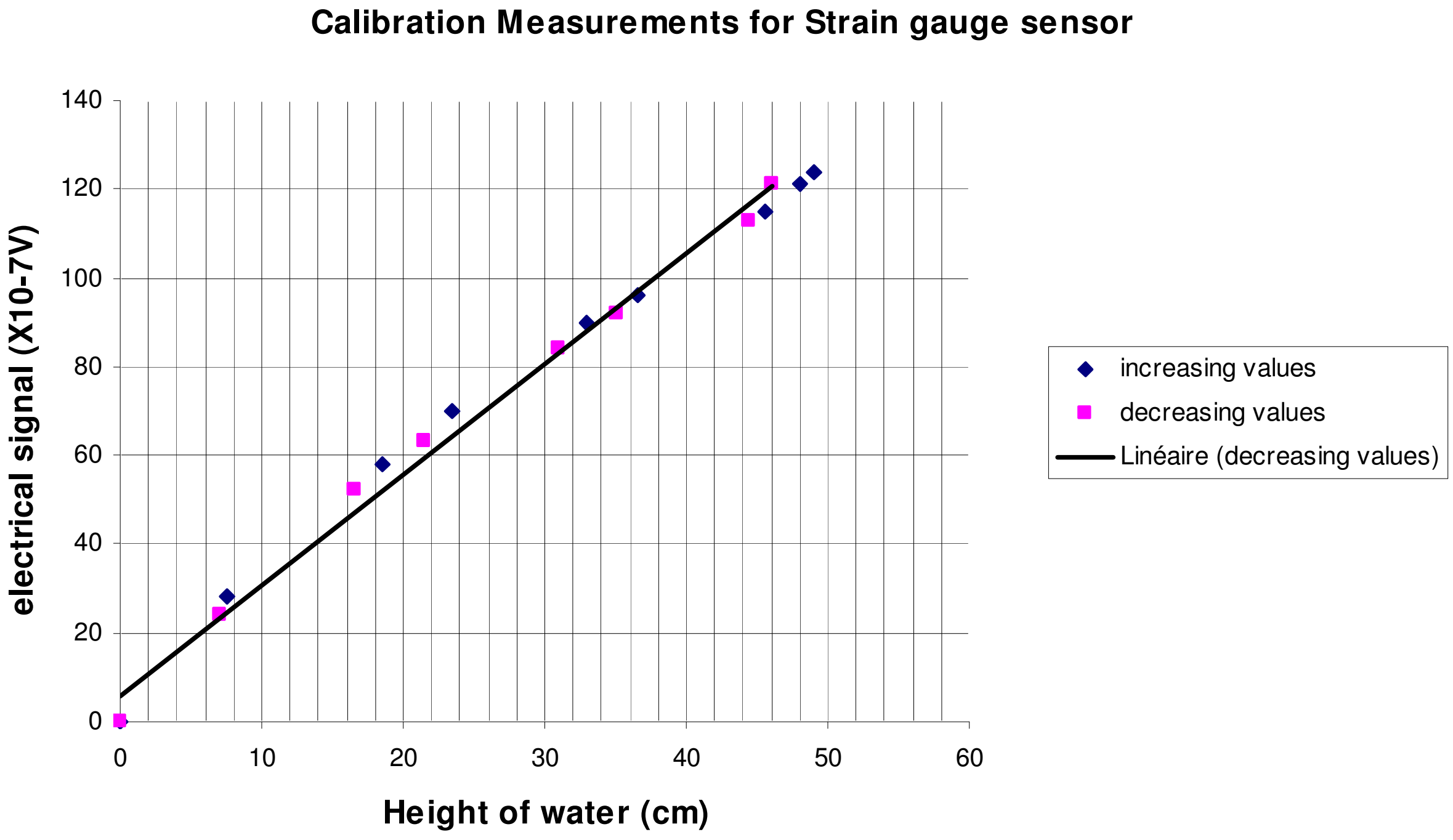

**Figure 7**

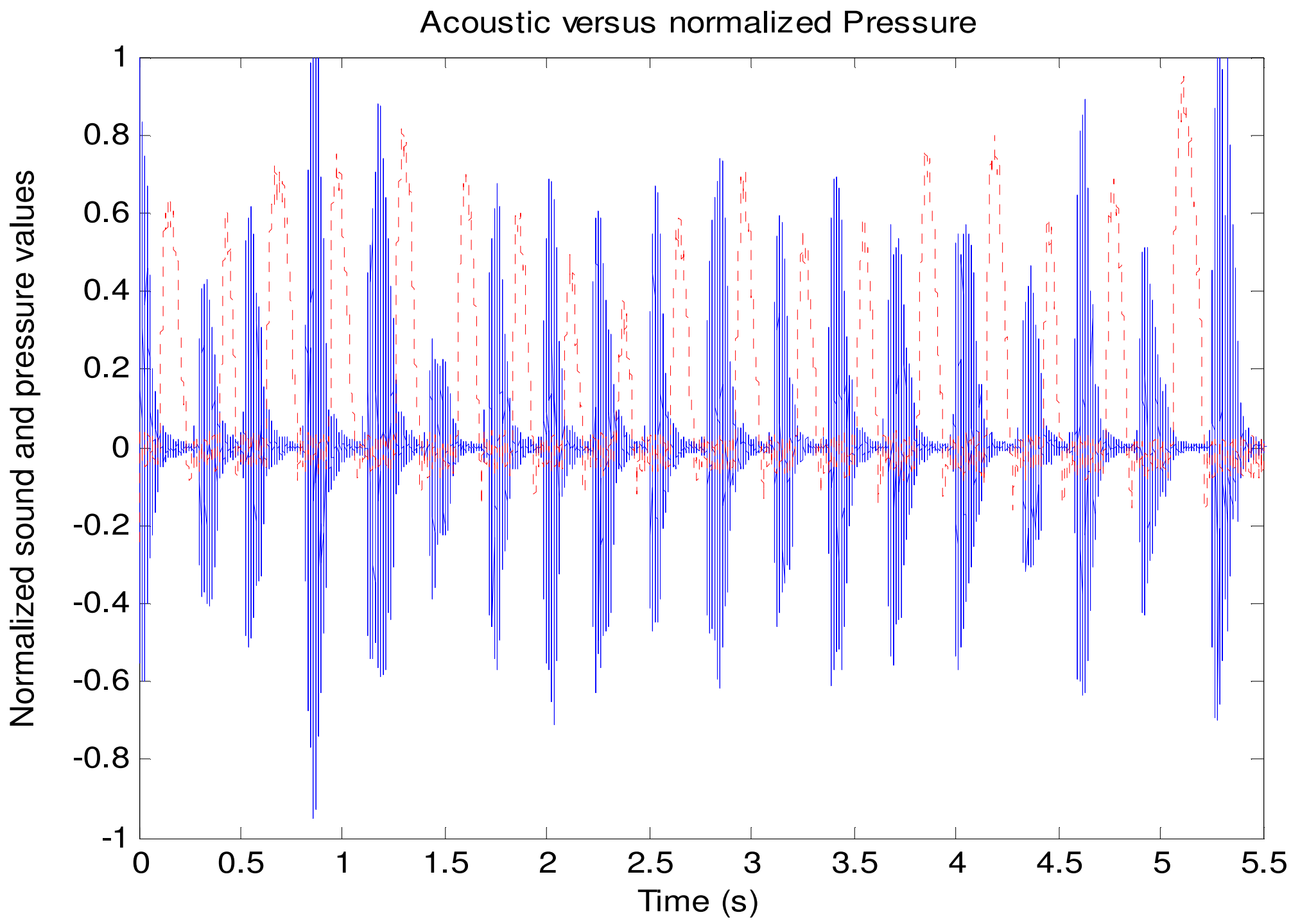

**Figure 8**

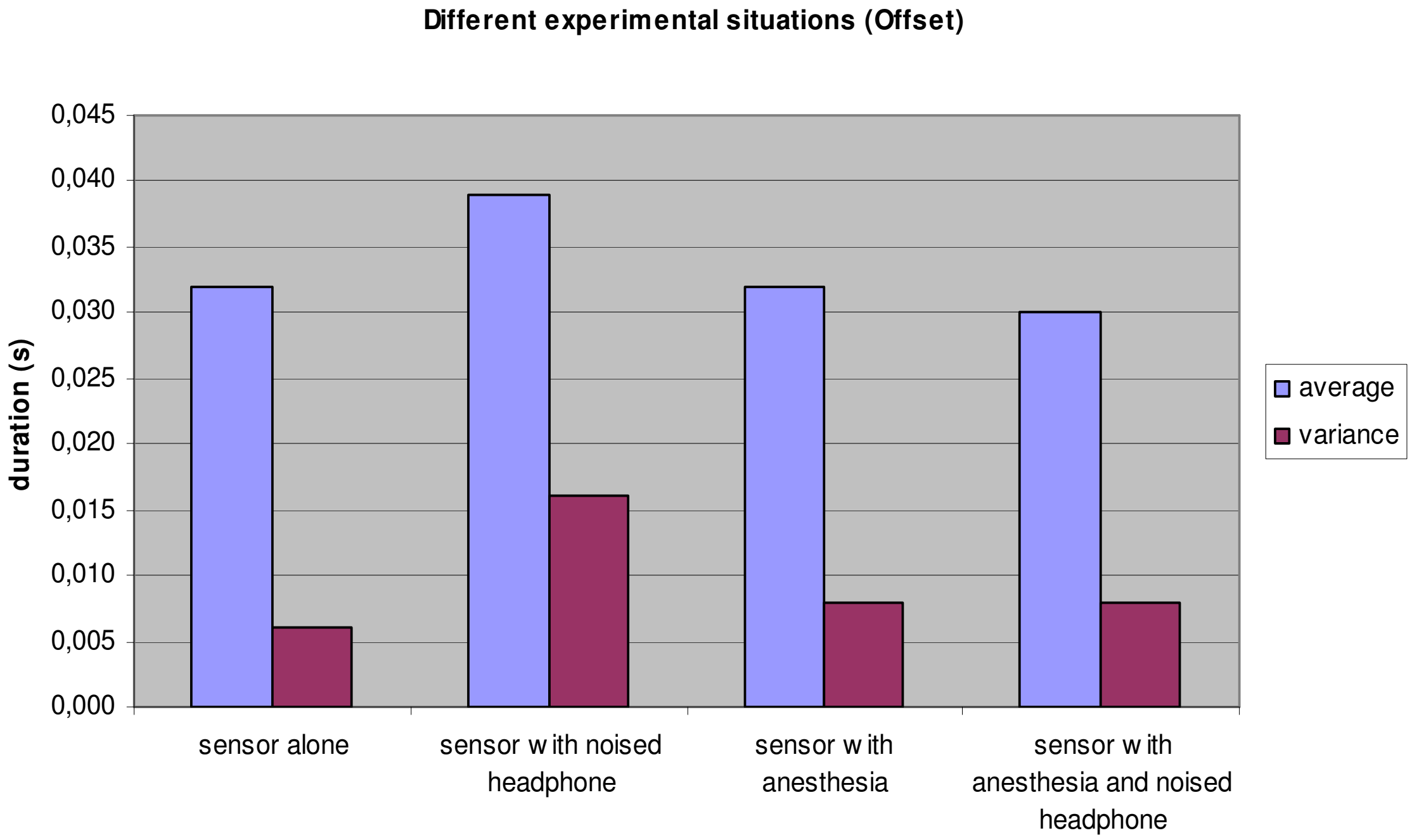

**Figure 9**

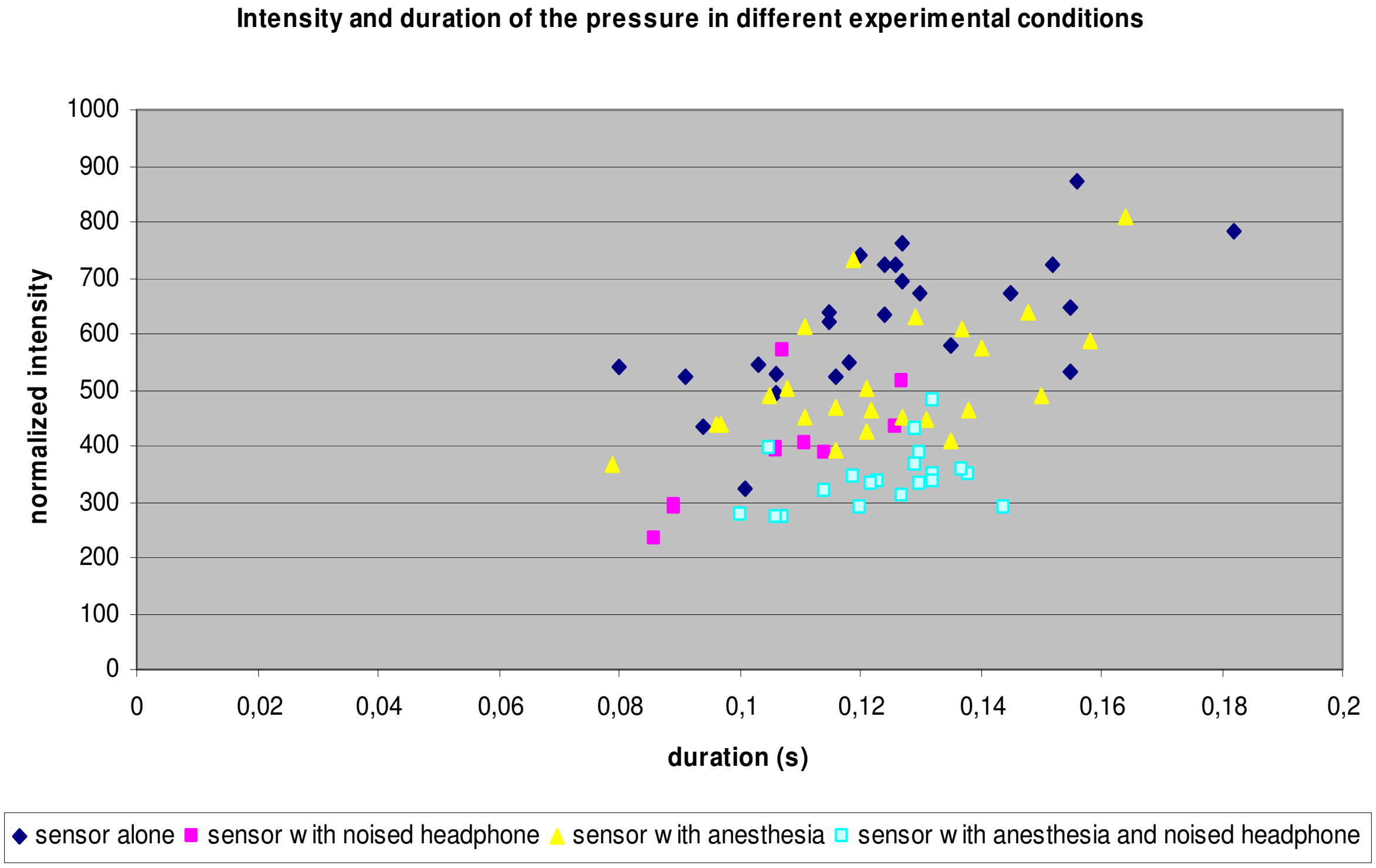

**Figure 10**